# Using Query Frequencies in Tree-Based Revocation for Certificateless Authentication in VANETs

Pino Caballero-Gil, Francisco Martín-Fernández and Cándido Caballero-Gil
Department of Computer Engineering
University of La Laguna
La Laguna, Spain
{pcaballe, francisco.martin.07, ccabgil}@ull.edu.es

*Abstract*—**Revocation of dishonest users is not an easy problem. This paper proposes a new way to manage revocation of pseudonyms in vehicular ad-hoc networks when using identity-based authentication to increase efficiency and security through certificateless authentication. In order to improve the performance of revocation lists, this paper proposes the use of a data structure based on authenticated dynamic hash k-ary trees and the frequency with which revoked pseudonyms are consulted. The use of the knowledge about the frequency of consultation of revoked pseudonyms allows an easier access to the most popular revoked pseudonyms to the detriment of revoked pseudonyms that are the least consulted. Accordingly, the proposal is especially useful in urban environments where there are vehicles that spend more time on road than others, such as public service vehicles.**

## I. INTRODUCTION

One of the main aspects to safeguard communication networks is authentication. Effective mechanisms are needed not only to authenticate legitimate and honest nodes, but also to have the ability to exclude malicious nodes in order to ensure network reliability.

When communication security is based on public-key cryptography, a central problem is to guarantee that a particular public-key is authentic and valid. The traditional approach to this problem is through public-key certificates emitted by a Public Key Infrastructure (PKI), in which a Certificate Authority (CA) certifies ownership and validity of public-key certificates. This solution presents many difficulties because the issues associated with certificate management are quite complicated and expensive. A different approach where the need for public-key certificates is deleted is the so-called Identity Based Cryptography (IBC), where each user's public key is his/her public IDentity (ID).

Efficient revocation is necessary when using public-key cryptography because private keys may become compromised. This problem has been traditionally solved through a centralized approach based on the existence of a Trusted Third Party (TTP), which is usually a CA distributing the so-called Certificate Revocation Lists (CRLs) that can be seen as blacklists of revoked certificates. Alternatively, some authors have proposed a different approach based on hash trees as Authenticated Data Structures (ADSs) in order to achieve a more efficient management of certificate revocation.

Vehicular Ad-hoc NETworks (VANETs) have been proposed as a solution to prevent adverse circumstances on the roads, and to achieve more efficient traffic management. They can be seen as self-organizing networks built up from moving vehicles that communicate with each other. In particular, these networks are considered an emerging research area of mobile communications because they offer a wide variety of possible applications, ranging from as aforementioned road safety and transport efficiency, to commercial services, passenger comfort, and infotainment delivery. Furthermore, VANETs can be seen as an extension of mobile ad-hoc networks where there are not only mobile nodes, named On-Board Units (OBUs), but also static nodes, named Road-Side Units (RSUs). The so-called Intelligent Transportation System (ITS) includes two types of communications: between OBUs and between OBUs, and RSUs [5]. Both the European standard for ITS, named ITS-G5, and its American counterpart, named Wireless Access in Vehicular Environment (WAVE), are based on the IEEE 802.11p amendment to the IEEE 802.11 standard which adds a vehicular communication system.

The open broadcasting of wireless communications and the high-speed mobility of vehicles in VANETs makes it more difficult to protect the security of communications. In these networks, any malicious misbehaving user that can inject false information, or modify/replay any previously disseminated message, could be fatal to the others. Therefore, within the family of standards for vehicular communications IEEE 1609 based on the IEEE 802.11p, the standard 1609.2 deals in particular with the issues related to security services for applications and management messages. This standard describes the use of PKIs, CAs and CRLs, and implies that in order to revoke a vehicle, a CRL has to be issued by the CA to the RSUs, who are in charge of sending this information to the OBUs. In particular, the IEEE 1609.2 standard proposes both broadcast authentication and non-repudiation through the use of the Elliptic Curve Digital Signature Algorithm (ECDSA). However, verifying every signature using ECDSA produces a high computational overhead.

On the one hand, according to these standards, each vehicle is assumed to have a pair of keys: a private signing key and a public verification key certified by the CA; and any VANET message must contain: a timestamp with the creation time, the sender's signature, and the sender's public-key certificate.

On the other hand, the so-called Dedicated Short-Range Communications (DSRC) channels specifically designed for automotive use, define that vehicles periodically exchange with nearby vehicles beacons containing sender's information such as location and speed because many VANET applications, such as the cooperative collision warning, rely on the information embedded in these beacons.

In order to protect privacy in VANETs, each OBU can obtain multiple certified key pairs and use different public keys each time. These public keys are linked to pseudonyms that allow preventing location tracking by eavesdroppers. Therefore, once VANETs are implemented in practice on a large scale, their size will grow rapidly due to the increasing number of OBUs and to the use of such multiple pseudonyms. Thus, it is foreseeable that if CRLs are used, they will grow up to become very large and unmanageable. Moreover, this context can bring a phenomenon known as implosion request, consisting of many nodes who synchronously try to download the CRL during its updating, producing serious congestion and overload of the network, which could lead to a longer latency in the process of validating a certificate.

This work defines the use of IBC to achieve certificateless and cooperative authentication in VANETs. It also introduces a Huffman k-ary hash tree as an ADS for the management of pseudonym revocation. By using this ADS, the process of query on the validity of public pseudonyms will be more efficient because OBUs will send queries to RSUs, who will answer them on behalf of the TTP. In this way, at the same time this TTP will no longer be a bottleneck and OBUs will not have to download any entire revocation list. Instead of that, they will have to manage hash trees where the leaf nodes contain revoked pseudonyms. In particular, the use of Huffman k-ary trees allows taking advantage of the efficiency of the Huffman algorithm to provide the shortest paths for the revoked pseudonyms that are more queried by network users.

This paper is organized as follows. Section 2 presents a review of related work. Concepts and notation used in the proposed authentication scheme based on the combination of IBC and Huffman k-ary hash trees are introduced in Section 3. Section 4 summarizes the main ideas of the proposal. Finally, Section 5 discusses some conclusions and open problems.

## II. RELATED WORK

In many cases, the use of public-key cryptography is essential for information security [2]. For instance, an aforementioned, the family od standards IEEE 1609 describes the use of PKI in VANETs. In particular, the work [9] describes a proposal for the use of a PKI to protect messages and mutually authenticate entities in VANETs. As a continuation of that work, the paper [19] defines a PKI-based security protocol where each vehicle preloads anonymous public/private keys and a TTP stores all the anonymous certificates of all the vehicles, but this scheme introduces inefficiency in the certificate management process.

Also based on a PKI, a well-known solution for strong authentication in VANETs is based on the signature of each message [11]. However, the use of a traditional approach to PKIs may fail to satisfy the real time requirement in vehicular communications because according to the DSRC protocol, each OBU will periodically transmit beacons so even in a normal traffic scenario, it is a very rigorous requirement to deploy an authentication scheme that allows at the same time efficient revocation of invalid public keys, and efficient use of valid public keys. This is exactly the main goal of this work.

A revocation method called Online Certificate Status Protocol (OCSP) involves a multitude of validation agents that respond to client queries with signed replies indicating the current status of a target certificate. This explicit revocation method has an unpleasant side effect because it divulges too much information. Since validation agents constitute a global service, they must involve enough replication to handle the load of all validation queries, what means that the signature key must be replicated across many servers, which is either insecure or expensive.

A solution called Certificate Revocation Tree (CRT) was proposed in [15] as an improvement for OCSP involving a single highly secure entity that periodically posts a signed CRL like data structure to many insecure validation agents so that users query these agents. In CRTs, the leaf nodes are statements concerning revoked certificates, and the CA signs the root. By using CRTs, the responder can prove the status of any certificate by showing the path from the root to the leaf node without signing the response, because the signatures of any leaf node are identical, and given by the signature contained in the root. Thus, no trust in the responder is necessary. The proposal here described is based on this idea, but does not use any certificate.

The basic ADS proposed in [15] is a Merkle hash tree [16] where the leaf nodes represent revoked certificates sorted by serial number. A client sends a query to the nearest agent, which produces a short proof that the target certificate is (or not) on the CRT. [12] introduces several methods to traverse Merkle trees allowing time space trade-offs. Other ADSs based on multi-dimensional tree structures are studied in [17] to support efficient search queries, allowing the retrieval of authenticated certificates from an untrusted repository used for dissemination by various credential issuers. Besides, many tree-balancing algorithms have been proposed in the bibliography for hash trees [4]. For instance, AVL trees are balanced by applying rotation, B-trees are balanced by manipulating the degrees of the nodes, and 2-3 trees contain only nodes with at least 2 and at most 3 children. However, in the particular application of public-key revocation, balancing trees does not necessary minimize the overall communication.

Another interesting problem with CRTs appears each time a certificate is revoked as the whole tree must be recomputed and restructured. Skip-lists proposed in [7] [8] can be seen as a natural and efficient structure to reduce communication by balancing the CRT. However, they are not good solutions for other problems such as insertion of new leaf nodes.

Hash trees are usually based on widely used hash functions. This work uses SHA-3 [1], which is a cryptographic hash function recently selected as the winner of the NIST hash function competition. SHA-3 uses the Keccak function and a sponge construction [1] in which message blocks are XORed into the initial bits of the state.

Finally, in order to solve the problem caused by the management of valid public-key certificates, the work [20] proposes the idea of an identity-based cryptosystem in which arbitrary strings can act as public keys so that there is no need for public-key certificates. The first practical identity-based encryption scheme was described in [3] using a bilinear map. Weil and Tate pairings on elliptic curves are the most efficient ways of constructing such bilinear maps [14]. The proposal here described was implemented using the Tate pairing for identity-based authentication.

## III. PRELIMINARIES

### A. ID-Based Cryptography

The idea of IBC and, in particular, of Identity-Based Signature (IBS) based on that the public identity *ID* of the signer can be used as verification key of a received signature, what avoids the need of any public-key certificate. In our scheme, such an identity is a public pseudonym $P_j$ sent by the signer node together with the signed message. In the used ID-based system, each node has to receive all the signing private keys $P_rP_j$ linked to all its pseudonyms $P_j$ from a TTP, because it cannot generate them by itself. In particular, a TTP, called in IBC the Private Key Generator (PKG), is in charge of computing and delivering to each node via a confidential channel, the signing private keys linked to each of its pseudonyms. On the other hand, the PKG publishes a master public key $MP_u$ and retains the corresponding master private key $MP_r$. Thus, given the master public key $MP_u$, any party will be able to compute the public key $P_uP_j$ corresponding to any pseudonym $P_j$ by combining it with $MP_u$. In order to use the corresponding private key, the node authorized to use a pseudonym must have received it from the PKG, which uses the master private key $MP_r$ to generate all the private keys corresponding to all the pseudonyms. Thus, the main algorithms in the proposed IBS are as follows:

- **Setup**: The PKG randomly picks its master private key $MP_r$, and therefore computes and publishes its master public key $MP_u$.

- **Extraction**: For each pseudonym $P_j$, the PKG uses its master private key $MP_u$ to compute the corresponding private key $P_rP_j$ and all pairs $(P_j, P_rP_j)$ are sent securely from the PKG to the corresponding owner.

- **Signature**: A signer node uses its private key $P_rP_j$ to compute the signature of a message $M$, and sends openly both the computed signature $PrP_j(M)$ and its pseudonym $P_j$.

- **Verification**: A node that receives a signed message and corresponding pseudonym $(P_rP_j(M), P_j)$ uses $MP_u$ and $P_j$ to compute $P_uP_j$ and verify the signature $P_rP_j(M)$.

Note that no new ID-based cryptosystem is described in this paper because it is out of its scope. The ID-based system that has been implemented in the proof of concept prototype is the Boneh-Franklin scheme [3], which uses a bilinear pairing over elliptic curves and bases its security on the Bilinear Diffie-Hellman problem.

The ID-based system is built from a bilinear map $e: G1 \times G1 \rightarrow G2$ between two groups $G1$ and $G2$ so that according to the bilinearity of $e: e(aP, bQ) = e(P,Q)ab$ for all $P, Q \in G1$ and $a,b \in Z$. Specifically, an ID-based system can be built from a bilinear map e if and only if a variant of the Computational Diffie-Hellman problem in $G1$ is hard. The considered Bilinear Diffie-Hellman problem in $G1$ is defined as follows: Given $P, aP, bP, cP$, compute $e(P,Q)abc$, where $P \in G1$ and $a, b, c \in Z$. In particular, the used bilinear pairing e is described for an elliptic curve $E$ defined over some field $K$, so it maps a pair of points of $E$ to an element of the multiplicative group of a finite extension of $K$.

The first satisfactory version of the Boneh-Franklin scheme was based on the Weil pairing [3]. However, the scheme implemented in this work uses the Tate pairing because this is considered the most convenient bilinear function for the Boneh-Franklin scheme in terms of computational cost. In particular, the implementation of the proposal includes the use of Miller's algorithm to compute the Tate pairing [17].

In IBC, just a few works exist on revocation mechanisms. However, proposals of new solutions to provide efficient mechanisms for key revocation are necessary. Here we propose a scheme to manage revoked pseudonyms in IBC-based VANETs, built on the idea of revocation hash trees.

### B. Tree Notation

The tree-based model described in this paper is based on the following notation:

- $h(.)$: Hash function used to define the revocation tree.

- $h(A0 \mid A1 \mid ...)$: Digest obtained with the hash function $h$ applied on the concatenation of the inputs $A_i, i=0, 1...$

- $D (\geq 1)$: Depth of the hash tree.

- $t$: total number of revoked pseudonyms.

- $RP_j (j=1, 2,..., t)$: $j$-th Revoked Pseudonym.

- $N_0$: Root Node of the hash tree.

- $N_{path}$: Node of the hash tree where *path* indicates branches that have to be taken to go from root node to leaf node.

- $k$: Maximum number of children for each internal node.

- $f(...)$: Keccak function used in SHA-3.

- $n$: Bit size of the digest of $h$.

- $s$: Bit size of the input to $f$.

- $r$: Bit size of the input blocks for $h$ after padding.

- $l$: Bit size of the output blocks that build the digest of $h$, which is here assumed to be lower than $r$.

### C. Huffman k-ary Hash Tree

. In order to improve efficiency of communication and computation in the management of revocations in VANETs, some authors have proposed the use of particular ADSs such as Merkle trees [6], Huffman Merkle trees [18] and skip lists [13]. However, to the best of our knowledge no previous work has described in detail the use of Huffman $k$-ary trees as hash trees for revoked pseudonym management.

In general, a hash tree is a tree structure whose nodes contain digests that can be used to verify larger pieces of data. Leaf nodes in a hash tree are hashes of data blocks while nodes further up in the tree are the hashes of their respective children so that the root of the tree is the digest representing the whole structure. Most implemented hash trees require the use of a cryptographic hash function $h$ in order to prevent collisions.

Like most hash trees, the Merkle tree is a binary tree, so each internal node $N_{ij}$ is the hash value of the concatenation of its two children: $N_{ij}=h(N_{i-1},0 / N_{i-1},1)$, where $i$ is the depth of node in the tree.

On the contrary, this work proposes the use of a more general structure known as k-ary tree, which is a rooted tree in which each node has no more than $k$ children, and each internal node is obtained by hashing the concatenation of all the digests contained in its children. Specifically, we propose the use of a Huffman $k$-ary tree in which leaf nodes are ordered from left to right, based on which revoked pseudonyms the most queried. Thus, we propose the introduction of the combination of both concepts of Huffman coding and $k$-ary trees applied to trees based revocation.

Huffman coding [10] is an algorithm used for data compression. The term refers to the use of a table of variable length codes for encoding certain symbols, where the table has been filled in a specific manner based on the best probability estimated of occurrence of each possible value of the symbol. Our proposal is to bring that idea to the CRLs, and allocate shortest paths in the tree to the revoked pseudonyms that are most queried. Thus, the tree will be built according to the consulted frequency in which the pseudonyms are by users of VANETs. In this way, the query of the consulting most revoked pseudonyms will be which more efficient.

Each node $N_{path}$ of our Huffman hash tree is given by a hash value. For each node $N_{path}$, $path$ is defined by the path from the root $N_0$ to the node $N_{path}$ (see Figure 1). The length of the path, given by the number of levels in the tree used in it, is related to the number of hash applications that are needed to represent the leaf node that corresponds to a revoked certificate.

The authenticity of the used hash tree structure is guaranteed thanks to the TTP signature of the root $N_0$. When a RSU answers to an OBU about a query on a pseudonym, it proceeds in the following way. If it finds the digest of the pseudonym among the leaf nodes of the tree, which means that it is a revoked pseudonym, the RSU sends to the OBU the route between the root and the corresponding leaf node, along with all the siblings of the nodes on this path. After checking all the digests corresponding to the received path, and the TTP signature of the root, the OBU gets convinced of the validity of the received evidence on the revoked pseudonym. Conceptually, thanks to the proposal of using a Huffman tree, queries regarding the most usually queried certificates involve less data transmission and computation.

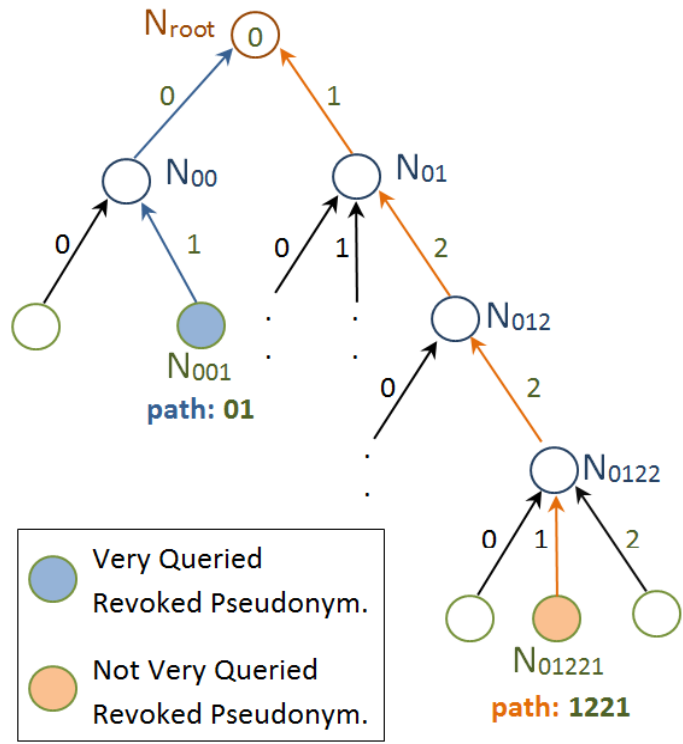

**Fig. 1. Revocation Pseudonym Path on Huffman 3-ary Tree**

### D. Construction of Huffman k-ary Tree

The idea of using the Huffman algorithm applied to CRLs is intended to optimize the query time and computation about revoked pseudonyms in VANETs. Generally, vehicles that spend more time on the roads are those that are more likely to communicate with other vehicles. Typical systems based on CRLs do not take into account this factor, so the average cost of finding any revoked pseudonym is the same. However, the general cost can be optimized by assigning less deep positions in the hash tree, to the most queried pseudonyms corresponding to the vehicles that stay longer on the road.

For this reason, the proposal includes the assumption that the RSU counts the number of queries that are performed for each revoked pseudonym. During the update of the tree, nodes are rearranged based on the new frequencies. Furthermore, taking into account the type of vehicle we can see that public vehicles (buses, taxis, etc.) are more likely to be among the most queried ones because they spend much time on the road.

In order to build the tree, the first factor to consider is the maximum and minimum number of children per node. This parameter defines the $k$ of the $k$-ary tree to be built. If this $k$ is equal to $2$, we get the typical binary Huffman tree. The proposed system allows other values for $k$, such as $3$, $4$, $5$, etc. Thus, if we propose a $k$-ary tree with a maximum of 5 children per node, we have a Huffman 5-ary like the one in Figure 2.

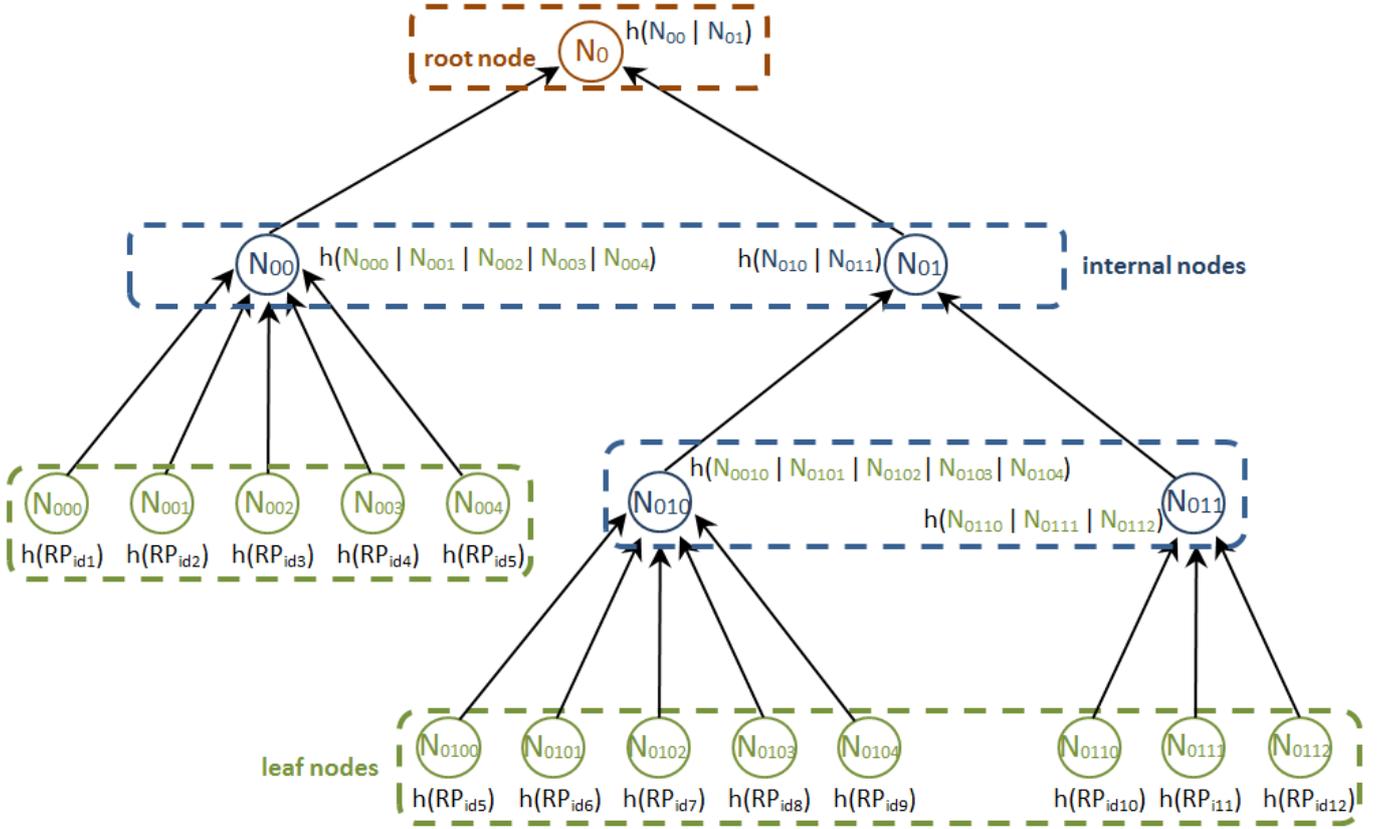

**Fig. 2. Hash Tree Based on Huffman 5-ary Tree.**

Once we know how many children per node are allowed as maximum, what we do is to create the Huffman tree so that internal nodes are assigned from the query frequencies calculated by the RSU. Whenever $k$ revoked pseudonyms are grouped in an internal node, this node is created with the sum of the frequencies of its children. In this way, the tree is constructed by grouping, first all revoked pseudonyms that are less consulted on new internal nodes, and so on to leave the most queried revoked pseudonyms in top positions in the tree. Thus, the search of these nodes is much faster and the route to the root node is much shorter.

In order to learn how to find a node in the tree, a hash table is used to map each revoked pseudonym with the exact path that defines the tree. Thus, if for example we have a Huffman 3-ary tree, we have that a node pseudonym $N_{01221}$ has a path in the tree *[1,2,2,1]* (see Figure 1), what means that from the root node we have to go through the branches starting by the branch *1*, then the *2* ... and so on to the internal node that is linked to it and opt for the branch *1* to get to it.

## IV. CERTIFICATELESS AUTHENTICATION

In the scheme proposed in this work, a node does not need any certificate to prove the binding to its public key. Instead of that, an ID-based authentication scheme and revocation trees are used. We consider the following basic authentication architecture, which includes three main parties:

- *TTP*: This entity acts as key distribution center because it is responsible for generating and assigning related parameters for VANET nodes, and for revoking pseudonyms of misbehaving OBUs and public keys of misbehaving RSUs.

- *RSU*: This entity serves as a gateway to provide OBUs within its transmission range with any requested information about revoked pseudonyms.

- *OBU*: Each vehicle is equipped with an OBU, which periodically broadcasts signed beacons that are received by neighbor OBUs and RSUs.

The proposed model is based on the use of a pseudonym $P_j$ set by each OBU, so that for each one the TTP provides the OBU with a corresponding private key $P_rP_r$. If any of those pseudonyms is revoked by the TTP, it inserts all the pseudonyms corresponding to the same OBU in the revocation tree. The TTP is also responsible for periodically updating the tree by deleting the expired pseudonyms, and for restructuring the tree when necessary. After each update, the TTP sends the corresponding modifications of the updated tree to all RSUs.

The RSU has to search vehicle pseudonyms in the revocation tree each time an OBU requests it. The RSU must provide the requesting OBU either with a verifiable revocation proof of any revoked pseudonym or with a signed message indicating that the requested pseudonym has not been revoked and is labelled as *OK*. In the first case, by using the answer data, the OBU can verify the TTP signature of the received signed root, recompute the root of the revocation tree, and check it by comparing it with the received signed root.

The proposed scheme is computationally efficient since it obviates the need to sign each RSU reply, as it removes most of the trust from it. The only case when the RSU's trust is questioned is when it provides an 'OK' answer because that could be a fraud.

In this regard, when an OBU receives an 'OK' message signed by a cheating RSU, it trusts it momentarily. However, when it contacts another RSU, it asks it again about the same pseudonym. If this RSU provides the OBU with a proof of revocation whose timestamp contradicts the 'OK' answer signed by the questioned RSU, the OBU sends to the latter RSU an impeachment on the questioned RSU, so that the honest RSU can send it to the TTP who will revoke its public key by deleting it directly from the revoked RSU. Otherwise, if the second RSU also sends a signed *OK* message, the OBU goes on asking about the same pseudonym until it reaches either a contradiction or a prefixed trust threshold.

Thus, each OBU stores locally in two separate and complementary structures, the pseudonyms of those OBUs that it has previously checked as unreliable, and of those OBUs that have been reliable till then. Therefore, in the future, if it reconnects with any of these vehicles, it can use such information to decide how to proceed. If there is no RSU nearby, it uses these data to decide whether to establish the communication or not. Otherwise, even if there is an RSU nearby, there is no need to re-ask it about a checked revoked pseudonym.

## V. Conclusions

Authentication is one of the most difficult security issues in vehicular ad-hoc networks. In particular, revocation of dishonest users is one the hardest problems. Because of this, identity-based cryptography is here proposed to achieve certificateless authentication, which increases efficiency and security of communications in VANETs. Moreover, to improve revocation management in VANETs, this paper introduces a new feature using a data structure based on authenticated dynamic hash k-ary trees combined with Huffman coding. This structure allows using a tree where the most queried revoked pseudonyms have a route to the root node shorter than the path of other pseudonyms.

This is part of a work in progress so there are still some open questions, such as the analysis of the optimal values of the parameters of the used k-ary trees, the study of the properties of the resulting Huffman code, and a comparison with previous proposals.

## Acknowledgment

Research supported by the MINECO of Spain and the FEDER Fund under Projects TIN2011-25452, IPT-2012-0585-370000 and the FPI scholarship BES-2012-051817.